\documentclass[twocolumn,showpacs,preprintnumbers,amsmath,amssymb]{revtex4}
\usepackage{graphicx}
\usepackage{dcolumn}
\usepackage{bm}
\usepackage{multirow}
\begin{document}


\title{Resonance excitations in $^7$Be(d,p)$^8$Be$^\ast$ to address the cosmological lithium problem}

\author{Sk M. Ali$^1$}
\email{mustak@jcbose.ac.in}
\author{D. Gupta$^1$}
\email{dhruba@jcbose.ac.in}
\author{K. Kundalia$^1$}
\author{Swapan K. Saha$^1$}		
\altaffiliation {Former faculty}
\author{O. Tengblad$^2$}
\author{J.D. Ovejas$^2$}
\author{A. Perea$^2$}
\author{I. Martel$^3$}		
\author{J. Cederkall$^4$}
\author{J. Park$^4$}
\altaffiliation{Present Address: Center for Exotic Nuclear Studies, Institute for Basic Science, 34126 Daejeon, South Korea}
\author{S. Szwec$^{5,6}$}
\affiliation{$^1$Department of Physics, Bose Institute, 93/1 APC Road, Kolkata 700009, India}
\affiliation{$^2$Instituto de Estructura de la Materia $-$ CSIC, Serrano 113 bis, ES-28006 Madrid, Spain}
\affiliation{$^3$University of Huelva, Av. Fuerzas Armadas s/n. Campus ``El Carmen", 21007, Huelva, Spain
}
\affiliation{$^4$Department of Physics, Lund University, Box 118, SE-221 00 Lund, Sweden}
\affiliation{$^5$Accelerator Laboratory, Department of Physics, University of Jyv$\ddot a$skyl$\ddot a$, FI-40014 Jyv$\ddot a$skyl$\ddot a$, Finland}
\affiliation{$^6$Helsinki Institute of Physics, University of Helsinki, FIN-00014 Helsinki, Finland}

\date{\today}
\begin{abstract}
The anomaly in the lithium abundance is a well-known unresolved problem in nuclear astrophysics. Recent revisit to the problem tried the avenue of resonance enhancement to account for the primordial $^7$Li abundance in standard Big-Bang Nucleosynthesis (BBN). Prior measurements of the  $^7$Be(d,p)$^8$Be* reaction could not account for the individual contributions of the different excited states involved, particularly at higher energies close to the Q-value of the reaction. We carried out an experiment at HIE-ISOLDE, CERN to study this reaction at E$_{cm}$ = 7.8 MeV, populating excitations up to 22 MeV in $^8$Be for the first time. The angular distributions of the several excited states have been measured and the contributions of the higher excited states in the total cross section at the relevant Big Bang energies were obtained by extrapolation to the Gamow window using the TALYS code. The results show that by including the contribution of the 16.63 MeV state, the maximum value of the total S-factor inside the Gamow window comes out to be 167 MeV b as compared to earlier estimate of 100 MeV b. However, this still does not account for the lithium discrepancy. 
\end{abstract}
\keywords{Resonance, Folding, Isospectral potential}
\pacs{25.40.Hs, 26.35.+c, 25.40.Ny, 25.60.-t, 29.30.Ep}
 
\maketitle

The cosmological lithium problem~\cite{CO04,BO10,FI11,CY16} is a decades-old and yet unresolved problem in nuclear astrophysics. The problem delineates a pronounced anomaly in abundance of $^7$Li between observation and prediction of the BBN theory. Using a baryon-to-photon ratio $\eta = (6.104 \pm 0.058) \times 10^{-10}$~\cite{AG18},  the BBN theory predicts a $^7$Li abundance about 3 times higher than observation. There is however, good agreement for $^2$H and $^4$He. Several avenues were searched unsuccessfully to solve the anomaly from observation errors or new physics beyond the standard model~\cite{FI11}. The primordial $^7$Li mostly originated from the decay of $^7$Be (T$_{1/2}$ = 53.22d~\cite{TI02}). Consequently the production and destruction channels of $^7$Be are linked to the lithium problem.

Previous experiments~\cite{AN05,RI19} on $^7$Be destruction with deuterium tried to search for a solution based on incomplete nuclear physics input to BBN calculations. The reaction $^7$Be(d,p)$^8$Be(2$\alpha$) has a high Q value of 16.67 MeV. The earliest cross-section measurements of this reaction by Kavanagh~\cite{KA60} were limited to excitations of 11 MeV. Later, Parker~\cite{PA72} estimated the contributions from higher excited states of $^8$Be, not observed in~\cite{KA60}. He multiplied the differential cross sections by 4$\pi$, and further by a factor of $\sim$ 3 to account for higher excited states. The d + $^7$Be rate used in BBN calculations~\cite{CA88} was based on~\cite{PA72}. Later work by Angulo~\cite{AN05} found even a smaller rate of d + $^7$Be. In addition to the ground state (gs) and first excited states of $^8$Be, the 11.4 MeV state in $^8$Be was observed and cross sections were measured up to excitations of 13.8 MeV~\cite{AN05}. It was concluded~\cite{AN05} that higher energy states not observed in~\cite{KA60} contribute $\sim$ 1/3 of the total S-factor instead of 2/3 estimated in~\cite{PA72}. 

Now, the destruction reaction might proceed through intermediate states; in $^8$Be by $^7$Be(d,p)$^8$Be$^*$; in $^5$Li by $^7$Be(d,$\alpha$)$^5$Li$^*$ or in a democratic three-particle decay of the $^9$B compound system~\cite{RI19}. Rijal stated in~\cite{RI19} that the (d,$\alpha$) yield dominated over (d,p) yield. They claim to have found a new resonance inside the Gamow window ($T = 0.5 - 1$ GK, $E_{cm} = 0.11 - 0.56$~MeV), that reduces the predicted abundance of $^7$Li but not sufficiently to solve it. However, Gai~\cite{GA19} pointed out that Rijal's new d + $^7$Be rate is nearly identical to~\cite{PA72} in the BBN region. Also the rates are uncertain by a factor of 10 due to uncertainty of the resonance energy around 16.8 MeV~\cite{GA19}. In this context, no data extracting contributions of the excited states around 16 MeV exist in the $^7$Be + d channel.

The above mentioned 16.8 MeV state of $^9$B , may also decay by proton emission to a highly excited state in $^8$Be at 16.626 MeV, subsequently breaking into two $\alpha$-particles~\cite{KI11}. To determine fully the contribution of $^7$Be(d,p)$^8$Be reaction to the $^7$Li abundance, it needs to be measured for $^8$Be excitations around 16 MeV. In order to measure that with good statistics, the experiment might be done at higher energies. The data are then used to normalize the excitation function calculated with TALYS~\cite{KO19}, to ascertain contributions of the higher excited states at the Gamow window.

In this Letter, we discuss our findings from the reaction $^7$Be(d,p)$^8$Be$^*$, measuring states of $^8$Be$^*$ up to 22 MeV. The experiment was carried out at CERN HIE-ISOLDE~\cite{HIE}, using a 5 MeV/u $^7$Be beam with energy spread of $\sim$ 168 keV. A uranium carbide target was irradiated offline~\cite{ST19} and the activated target was heated up during the experiment. Using the RILIS~\cite{RILIS} laser ion-source, the $^7$Be was extracted and post-accelerated. When the residual gas in the REX-EBIS charge breeder is ionized~\cite{REX}, the stable beam contaminants produced are mainly $^{14}$N$^{4+}$ and $^{21}$Ne$^{6+}$, having the same A/q as $^7$Be$^{2+}$. A stripping foil and a dipole before the experimental station was used to clean the beam to $^7$Be$^{4+}$. Any $^7$Li accelerated as $^7$Li$^{2+}$ was fully stripped in the foil and removed. The intensity of $^7$Be beam was $\sim$ 5 $\times$ 10$^5$ pps impinging on a 15 $\mu$m CD$_2$ target. We also used a 15 $\mu$m CH$_2$ and a 1 mg/cm$^2$ $^{208}$Pb target for background measurements and normalization respectively, by taking short consecutive runs with $^{208}$Pb and CD$_2$ targets.
\begin{figure}[h!]
\includegraphics[width=0.5\textwidth]{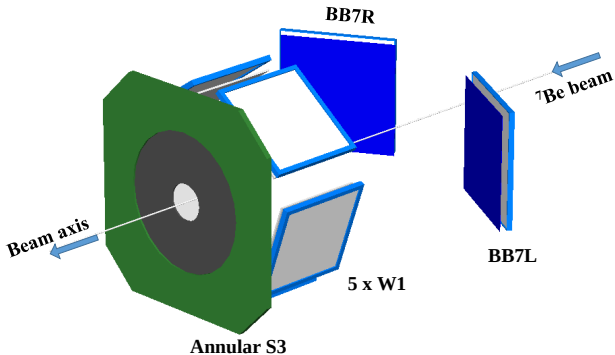}
\caption{\label{fig:setup} The detector setup for the $^7$Be + d experiment at 5~MeV/u.}
\end{figure}

The setup (Fig.~\ref{fig:setup}) in the scattering chamber~\cite{SEC}, consisted of 5 double-sided Micron 16$\times$16 silicon strip detectors (DSSD) of thickness 60 $\mu$m (W1). These were backed by unsegmented 1500 $\mu$m silicon pad detectors (MSX25). The $\Delta E-E$ telescopes were set up in a pentagon geometry covering 40$^{\circ}$ $-$ 80$^{\circ}$ in lab. Each W1 was symmetrically placed with respect to the target center at a distance of 59 mm. The angles 8$^{\circ}$ $-$ 25$^{\circ}$ were covered by a 1000 $\mu$m annular detector (S3), placed at a distance of 74.1 mm from the target. The angles 127$^{\circ}$ $-$ 165$^{\circ}$ were covered by two 32$\times$32 DSSDs of thickness 60 $\mu$m and 140 $\mu$m (BB7), placed right and left of the beam direction respectively. The distance of the left (right) BB7 are  102 (110) mm from the target. These were backed by unsegmented 1500 $\mu$m silicon pad detectors (MSX40). The total solid angle coverage of the detectors is $\sim$ 32$\%$ of 4$\pi$. 
\begin{figure}[h!]
\includegraphics[width=0.5\textwidth]{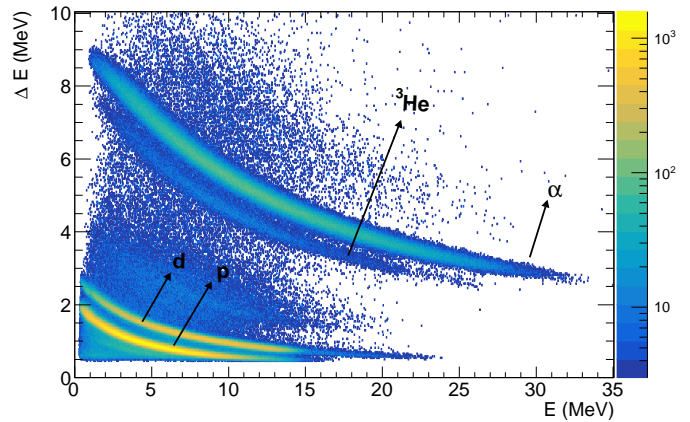}
\caption{\label{fig:dE-E} The $\Delta E-E$ spectrum of protons, deuterons, $^{3}$He and $\alpha$, detected at W1 + MSX25 telescopes, from 5 MeV/u $^7$Be on CD$_2$ target.}
\end{figure}

The detectors were calibrated with a $^{148}$Gd$-^{239}$Pu$-^{241}$Am$-^{244}$Cm mixed $\alpha$-source. For S3, due to higher dynamic range, we also used the elastic peaks from Rutherford scattering of ${5}$ MeV/u $^{7}$Be and ${5.15}$ MeV/u $^{12}$C beams on $^{208}$Pb. The light charged particles emitted from $^7$Be + d reaction, were identified by $\Delta E-E$ telescopes from the energy loss spectra (Fig. \ref{fig:dE-E}). Elastic $^7$Be scattered from Carbon of CD$_2$ was observed in S3.

\begin{figure}[h!]
\includegraphics[width=0.5\textwidth]{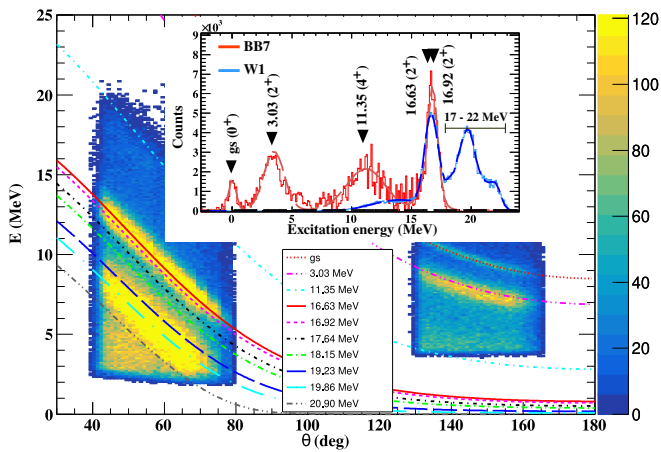}
\caption{\label{fig:E-Th} $E$ vs $\theta$ for protons at W1 and BB7. The kinematic lines for different excited states of $^8$Be are shown for $^7$Be(d,p)$^8$Be$^*$ at 5 MeV/u. Inset shows the excitation energy spectrum of $^8$Be.}
\end{figure}

Data from both W1 and BB7 were used to study excitations in $^8$Be. At BB7, we detected the gs and 3.03, 11.35, 16.63 MeV excited states of $^8$Be. After selecting the protons from $\Delta E-E$ spectra, the gs and 3.03 MeV could be identified from the lab energy ($E$) vs scattering angle ($\theta$) plots (Fig.\ref{fig:E-Th}). The protons corresponding to 11.35 and 16.63 MeV states were completely stopped in $\Delta E$ of BB7 and were extracted from the excitation energy spectrum (Fig. \ref{fig:E-Th} inset) after background subtraction. Since Fig. \ref{fig:E-Th} shows only protons selected from $\Delta E-E$ spectrum, we do not observe kinematic bands corresponding to these two states at $130^\circ-170^\circ$. The higher excited states above 16 MeV were obtained from W1 using $E$ vs $\theta$ plot (Fig. \ref{fig:E-Th}) of protons detected in coincidence with $\alpha$-particles. To obtain data for forward scattered protons, the energy-energy correlation of two coincident $\alpha$-particles at W1 with one hit at S3 were considered. The $\alpha$-p coincidence efficiency at W1 drops sharply for excitation energy below 16 MeV. Hence the kinematic signatures corresponding to states below 16 MeV are absent at $40^\circ-80^\circ$.
\begin{table}[h!]
\caption{\label{tab:table1}%
Excitations (E$_x$) of $^8$Be listed in Tilley et al~\cite{TI04}, and excitations (E$_x^{fit}$) and widths ($\Gamma^{fit}$) obtained from fitting the excitation energy spectrum in the inset of Fig. \ref{fig:E-Th}.}
\begin{ruledtabular}
\begin{tabular}{llcl}
\textrm{E$_x$ (MeV $\pm$ keV)}&
\textrm{$\Gamma$ (keV)}&
\textrm{E$_x^{fit}$ (MeV)}\footnote{The average uncertainty in the peak centroids is around $10\%$.}&
\textrm{$\Gamma^{fit}$ (MeV)}\\
\colrule 
0.0               & $5.57\pm 0.25$ eV     & 0.02   & $0.71 \pm 0.04$ \\
$3.03  \pm 10$    & $1513 \pm 15$         & 3.51   & $2.00 \pm 0.50$ \\
$11.35 \pm 150$   & 3500                  & 11.31  & $3.77 \pm 0.27$ \\
$16.626 \pm 3$    & $108.1 \pm 0.5$       & 16.50  & $1.18 \pm 0.51$ \\
$16.922 \pm 3$    & $74.0 \pm 0.4$        & 16.99  & $1.17 \pm 0.51$ \\
$17.640 \pm 1.0$  & $10.7 \pm 0.5$        & 17.50  & $0.71 \pm 0.05$ \\
$18.150 \pm 4$    & $138 \pm 6$           & 18.19  & $0.71 \pm 0.05$ \\
18.91             & 122                   & 18.80  & $0.67 \pm 0.01$ \\
$19.07 \pm 30$    & $270 \pm 20$          & 19.00  & $0.90 \pm 0.25$ \\
$19.235 \pm 10$   & $227 \pm 16$          & 19.28  & $0.71 \pm 0.03$ \\
19.40             & 645                   & 19.58  & $0.95 \pm 0.27$ \\
$19.86 \pm 50$    & $700 \pm 100$         & 19.79  & $0.95 \pm 0.25$ \\
20.10             & $880 \pm 20$          & 20.00  & $1.10 \pm 0.22$ \\
20.20             & $720 \pm 20$          & 20.29  & $1.41 \pm 0.69$ \\
20.90             & $1600 \pm 200$        & 20.84  & $1.75 \pm 0.15$ \\
21.50             & 1000                  & 21.44  & $1.60 \pm 0.60$ \\
$22.05 \pm 100$   & $270 \pm 70$          & 22.10  & $0.94 \pm 0.28$   \\
\end{tabular}
\end{ruledtabular}
\end{table}
The energy resolution $\sim$ 660 keV due to beam, target straggling and detectors, limits the separation of narrowly spaced high lying resonances at 16.63 and 16.92 MeV, and resonances around $17-22$ MeV. Earlier works~\cite{WI65,BE17} suggest that the 16.63 MeV state is populated considerably more than 16.92 MeV. Hence, we refer to this doublet as 16.63 MeV. The inset of Fig. \ref{fig:E-Th} shows excitation energy spectrum of $^8$Be from protons detected at W1 (blue) and BB7 (red). The corresponding fits are shown by dark-blue and brown dashed lines. The counts from BB7 are multiplied by two for clarity. The $16-22$ MeV excited states are populated for the first time from $^7$Be(d,p)$^8$Be$^*$. We did not find any evidence of (d,$\alpha$) channel. The peaks are fitted simultaneously by Gaussian functions, with input of excitation energies and their widths~\cite{TI04}. The arrows show the excitation energies used in the fitting. The peak centroids and widths from the resultant fits are summarized in Table~\ref{tab:table1}. The fitted widths represent the quadratic sum of 660 keV experimental resolution  and total decay width of the excited states.

\begin{figure}[h!]
\includegraphics[width=0.5\textwidth]{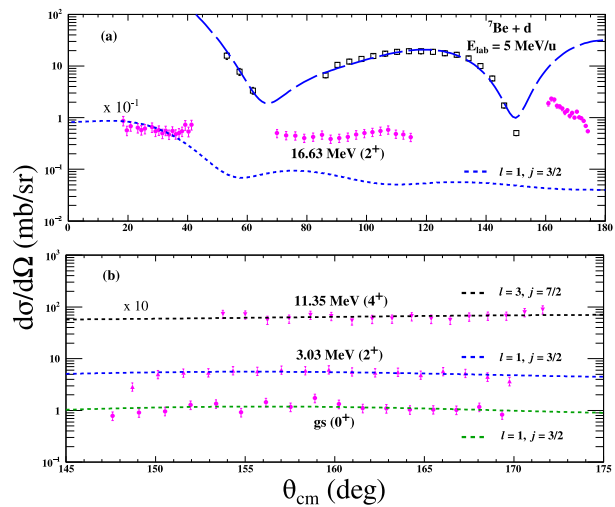}
\caption{\label{fig:dsig} Angular distributions for $^7$Be(d,d) and $^7$Be(d,p) to 16.63 MeV (a);  to 0.0, 3.03 and 11.35 MeV states (b). The corresponding  FRESCO~\cite{TH88} calculations are shown by dashed and dotted lines respectively.}
\end{figure}

The angular distribution for $^7$Be + d elastic scattering from the present work is shown in Fig. \ref{fig:dsig}(a). The scattering is quasi elastic as the 0.43 MeV state of $^7$Be could not be separated. However, the inelastic contribution is expected to be small~\cite{DA06}. The $^7$Be(d,p)$^8$Be* angular distributions for 16.63 MeV ($\times 10^{-1}$ for clarity) and gs, 3.03, 11.35 MeV ($\times 10$ for clarity) excited states are shown in Fig. \ref{fig:dsig}(a)(b). At BB7, corresponding to excited states up to 16.63 MeV, the proton counts are obtained from background subtracted excitation energy histogram with an angular interval of two degrees. For excited states $> 11.35$ MeV detected at W1, we considered the coincidence efficiency to correct for proton counts detected in coincidence with $\alpha$-particles using NPTool~\cite{MA16}. The errors in cross sections mainly arise from statistical uncertainties, systematic uncertainties in target thickness ($\sim 10\%$) and beam intensity ($\sim 10\%$). The optical model potential (OMP) fit for (d,d) and finite range DWBA calculations for (d,p) are carried out by FRESCO~\cite{TH88}, shown by dashed and dotted lines in Fig. \ref{fig:dsig}. The DWBA calculations require the OMP for the entrance channel d + $^7$Be, exit channel p + $^8$Be, and the core-core p + $^7$Be interactions. Woods-Saxon interaction potentials are used~\cite{PE76}.
\begin{table*}[th!]
\caption{\label{tab:OMP}Optical model parameters used in the present work. $V$ and $W$ are the real and imaginary depths in MeV, $r$ and $a$ are the radius and diffuseness in fm. $R_x = r_x A^{1/3}$ fm ($x=V, S, SO, C$).}
\begin{ruledtabular}
\begin{tabular}{c  c  c   c  c  c  c  c  c  c  c  }
Channel &$V$ &$r_{V}$ &$a_{V}$  &$W_{S}$ &$r_{S}$ &$a_{S}$ &$V_{SO}$ &$r_{SO}$ &$a_{SO}$ &$r_C$  \\
\hline 
d + $^7$Be &80.98 &1.35   &0.83     &36.91  &2.21   &0.10  &2.08    &0.49  &0.42 &1.30     \\
p + $^7$Be &92.07  &0.87   &0.89     &1.23  &0.10   &0.10  &16.82  &1.34  &0.12 &1.14   \\
p + $^8$Be   &82.60  &1.10   &0.41    &1.97  &1.10    &1.30    &5.54  &1.14  &0.57  &1.14  \\
\end{tabular}
\end{ruledtabular}
\end{table*}
\begin{figure}[h!]
\includegraphics[width=0.5\textwidth]{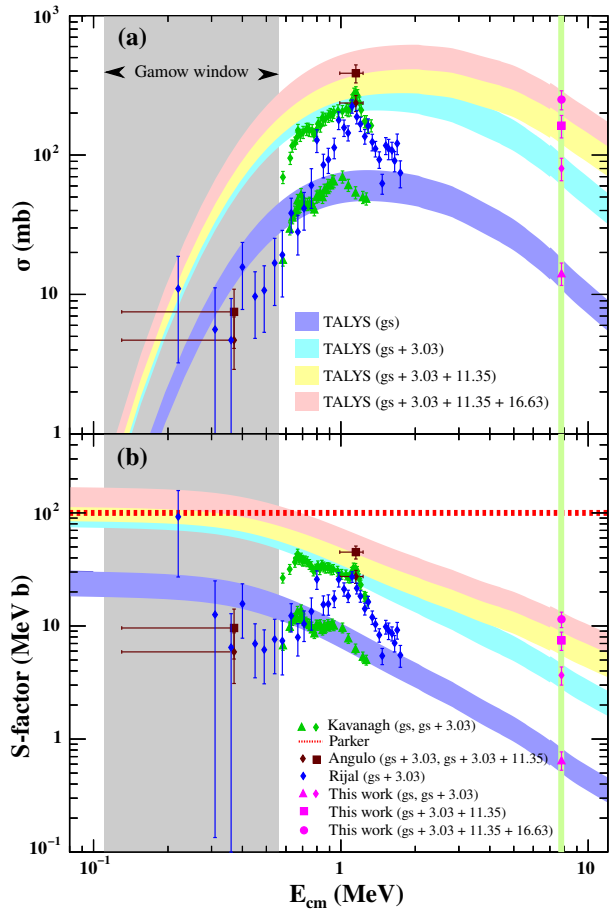}
\caption{\label{fig:TALYS}(a) Excitation function for $^7$Be(d,p)$^8$Be$^*$. The solid triangle, diamond, square and circles correspond to total cross-sections due to gs, gs + 3.03, gs + 3.03 + 11.35 and gs + 3.03 + 11.35 + 16.63 MeV states respectively. The data in green, brown, blue and magenta are the measurements of \cite{KA60,AN05,RI19} and the present work. The violet (gs), cyan (gs + 3.03), yellow (gs + 3.03 + 11.35) and red (gs + 3.03 + 11.35 + 16.63) MeV bands are TALYS calculations normalized to the present data at 7.8 MeV (green vertical line). The bands do not include systematic uncertainty due to extrapolation. (b) The S-factor representation of the excitation function. The red dotted line is the estimate by Parker~\cite{PA72}.}
\end{figure}
We started with OMP for d + $^7$Be~\cite{PA17}, p + $^7$Be~\cite{WA69} and used SFRESCO~\cite{TH88} to arrive at the final OMP (Table \ref{tab:OMP}) by minimizing $\chi^2$. Since p + $^8$Be elastic data are not available, we used OMP of p + $^8$Li~\cite{PA17} (Table \ref{tab:OMP}) for the exit channel. The n + $^7$Be binding potential~\cite{SC71} has a Wood-Saxon shape with $r=1.36$ fm and $a=0.55$ fm. Its depth is adjusted to reproduce effective neutron separation energy for each state of $^8$Be, while a spin-orbit potential with same geometry and fixed well depth of 9 MeV is included. A Gaussian interaction $V_{np} (r) = -V_0 \exp(-r^2 /r^2_0)$ is used for the n$-$p system with $V_0 = 72.15$ MeV and $r_0 = 1.484$ fm~\cite{AU87}. The DWBA calculations are in good agreement with the data for gs, 3.03 and 11.35 MeV states in a limited angular range while for the 16.63 MeV state, only forward angles up to $\sim 40^\circ$ could be reproduced. Consideration of compound nuclear contributions, coupled-channel calculations incorporating collective excitations of $^7$Be, inclusion of deuteron breakup and a d + $^7$Be adiabatic potential might result in a better fit at large angles. The experimental spectroscopic factors $C^2 S$, are extracted by normalizing the calculated DWBA cross-sections to measured cross-sections. The average values of $C^2 S$ for the excited states of $^8$Be are compared with the OXBASH~\cite{Br04} shell model calculations~\cite{Ku74} in Table ~\ref{tab:CS}. Assuming full isotropy for the gs, 3.03 and 11.35 MeV states, we obtain the total cross section $\sigma$ for these states. For 16.63 MeV, $\sigma$ is obtained by connecting the differential cross sections and integrating (Table \ref{tab:CS}).
\begin{table}[h!]
\caption{\label{tab:CS}The spectroscopic factors $C^2S$ from the present work as compared to the theoretical predictions of~\cite{Ku74}.}
\begin{ruledtabular}
\begin{tabular}{c c c c  c c c } 
E$_x$ (MeV)  &{T}  &{$nl_j$}  &{$C^2 S$} &{Kumar\cite{Ku74}} &{$\sigma$ (mb)}   \\ 
\hline                    
0.0     &0  &$1p_{3/2}$  &$1.19\pm 0.22$  &1.51     &$14.2 \pm 2.6$  \\
\\
3.03    &0  &$1p_{3/2}$  &$1.11\pm 0.20$  &0.91     &$65.9 \pm 12.1$ \\
\\
11.35   &0  &$1f_{7/2}$  &$3.20\pm 0.59$  &$-$         &$82.3 \pm 15.1$ \\
\\
16.63   &$0+1$ &$1p_{3/2}$ &$0.35\pm 0.07$ &0.34    &$87.5 \pm 9.2$   \\
\end{tabular}
\end{ruledtabular}
\end{table}
To obtain the excitation functions of $^7$Be(d,p)$^8$Be$^*$, TALYS-1.95~\cite{KO19} calculations are carried out and normalized to present experimental data.  In Fig. \ref{fig:TALYS}(a), the normalized excitation functions from TALYS for gs, gs + 3.03, gs + 3.03 + 11.35 and gs + 3.03 + 11.35 + 16.63 MeV (colored bands to include error bars of the present data) have been compared to the data of~\cite{KA60,AN05,RI19}. The calculations  agree very well with~\cite{KA60,AN05} outside the Gamow window, for E$_{cm}$ $>$ 0.7 MeV. The data~\cite{RI19} for gs and 3.03 MeV are well below corresponding TALYS calculations (cyan). However, the data within and near Gamow window have relatively large errors. Inside Gamow window, TALYS calculations for higher excited states (cyan, yellow, red) overestimate~\cite{AN05,RI19} except at 0.22 MeV, within error bars. The systematic uncertainties due to extrapolation of the total S-factor from E$_{cm}$ = 7.8 MeV to BBN energies is $\sim$ 48$\%$. This arises due to various phenomenological and microscopic models for level densities ($\sim$ 46$\%$), choice of global deuteron ($\sim$ 10$\%$) and proton ($\sim$ 10$\%$) OMP in TALYS. In Fig. \ref{fig:TALYS}(b), the corresponding astrophysical S-factor is shown. The S-factor band with contributions from gs + 3.03 + 11.35 MeV (yellow) converges to an average value of $\sim 95.6$ MeV b in the Gamow window which is close to Parker's estimate of 100 MeV b. The S-factor band (red) with the contributions of all the states including 16.63 MeV gives a maximum of 167 MeV b inside the Gamow window, close to 162.0 MeV b for the claimed new resonance at 0.36 MeV~\cite{RI19}. If we assume a constant S-factor of 167 MeV b, then the ratio of reaction rate from the present work with~\cite{CA88} at the relevant BBN energies is less than 2 whereas for solving the Li problem this ratio needs to be around 100~\cite{CO04}. Therefore, the BBN calculations of~\cite{RI19} seem to overestimate the importance of the $^7$Be + d reaction. Even the maximum S-factor inferred from this measurement would reduce the primordial Li abundance by less than 1$\%$~\cite{FI20,DA20} with respect to previous expectation~\cite{CA88} and thereby fail to alleviate the discrepancy between theory and observation. 

In summary, the present experiment reports the first measurement of all resonances in the $^7$Be(d,p)$^8$Be$^*$ channel up to 22 MeV. In particular, the 16.63 MeV state is analyzed in the context of the cosmological lithium problem. The measurement was carried out at a much higher center-of-mass energy of 7.8 MeV compared to the Gamow window. This facilitated populating the previously unseen higher excitations of $^8$Be with good statistics and their contributions to the total cross section are studied. The existing data within Gamow window has large error bars in energy as well as cross sections. The TALYS calculations normalized to the present data give an estimate of the contributions of resonance excitations in the (d,p) channel. It is apparent that inclusion of the 16.63 MeV state may lead to a maximum S-factor of 167 MeV b, higher than earlier used value of 100 MeV b~\cite{PA72}. But it does not fully account for the Lithium anomaly and our present understanding of nuclear physics may not solve this problem.

\begin{acknowledgments}
The authors thank the ISOLDE engineers in charge, RILIS team and Target Group at CERN for their support. D. Gupta acknowledges research funding from the European Union's Horizon 2020 research and innovation programme under grant agreement no. 654002 (ENSAR2) and ISRO, Government of India under grant no. ISRO/RES/2/378/15$-$16. O. Tengblad would like to acknowledge the support by the Spanish Funding Agency (AEI/FEDER, EU) under the project PID2019-104390GB-I00. I. Martel would like to acknowledge the support by the Ministry of Science, Innovation and Universities of Spain (Grant No. PGC2018-095640-B-I00). J. Cederkall acknowledges grants from the Swedish Research Council (VR) under contract numbers VR-2017-00637 and VR-2017-03986 as well as grants from the Royal Physiographical Society. J. Park would like to acknowledge the support by Institute for Basic Science (IBS-R031-D1). S. Szwec acknowledges support by the Academy of Finland (Grant No. 307685).

\end{acknowledgments}  


\end{document}